\documentclass[aps,prb,twocolumn,showpacs,superscriptaddress]{revtex4}
\usepackage{amssymb,graphicx,amsmath,enumerate}

\begin{document}

\title[Nanostructures in p-GaAs with improved tunability]{Nanostructures in p-GaAs with improved tunability}

\author{M. Csontos}
\author{Y. Komijani}
\affiliation{Solid State Physics Laboratory, ETH Zurich, 8093 Zurich, Switzerland}

\author{I. Shorubalko}
\affiliation{Solid State Physics Laboratory, ETH Zurich, 8093 Zurich, Switzerland}
\affiliation{Electronics/Metrology/Reliability Laboratory, EMPA, 8600 Duebendorf, Switzerland}

\author{K. Ensslin}
\affiliation{Solid State Physics Laboratory, ETH Zurich, 8093 Zurich, Switzerland}

\author{D. Reuter}
\author{A. D. Wieck}
\affiliation{Angewandte Festk\"orperphysik, Ruhr-Universit\"at Bochum, 44780 Bochum, Germany}

\date{\today}

\begin{abstract}
A nano-fabrication technique is presented which enables the fabrication of highly tunable devices on p-type, C-doped GaAs/AlGaAs heterostructures containing shallow two-dimensional hole systems. The high tunability of these structures is provided by the complementary electrostatic effects of intrinsic in-plane gates and evaporated metallic top-gates. Quantum point contacts fabricated with this technique were tested by electrical conductance spectroscopy.
\end{abstract}


\maketitle

The recently emerging interest in p-type nanodevices is two-fold. On one hand the pronounced spin-orbit as well as carrier-carrier Coulomb interactions make these systems suitable for studying many-body phenomena in reduced dimensions. On the other hand, low-dimensional hole-doped systems represent promising candidates for the practical implementation of spin-based quantum information processing technology.\cite{Wolf2001} Hyperfine interactions, identified as one of the main sources of spin dephasing in n-type systems,\cite{Khaetskii2002} are expected to be strongly suppressed due to the p-type nature of the valence band holes.\cite{Gerardot2008} Additionally, while strong spin-orbit interactions lead to fast spin relaxation in higher dimensions, further confinement of holes into quantum dots can significantly increase the relaxation time of the holes' spins,\cite{Bulaev2005} so that it can be comparable, or even larger than the one of the electron spins as it was tested by optical pumping experiments in self-assembled p-type quantum dots.\cite{Gerardot2008,Brunner2009}

In spite of the above prospects, low-dimensional hole systems have received less attention due to technological difficulties in fabricating stable p-type structures. Nanodevices defined on shallow two-dimensional hole gases (2DHGs) in p-type GaAs by conventional split-gate techniques show significant gate instabilities. This is presumably due to the fact that metallic Schottky barriers on p-type GaAs are leaky and exhibit a highly hysteretic behavior, making them unsuitable for high-precision tuning of the devices. In previous works\cite{Grbic2005,Grbic2007,Komijani2008} local anodic oxidation lithography\cite{Held1998} carried out by a charged tip of an atomic force microscope (AFM) has been utilized as an alternative method for establishing stable nano-constrictions tuned by intrinsic in-plane side-gates.\cite{Wieck1990}

However, the tunability of the nanostructures patterned on p-GaAs turns out to be much more sensitive to specific oxide line parameters compared to those implemented in n-GaAs resulting in a lower yield of the functional devices. In this paper we demonstrate that combining the electrostatic effect of such in-plane gates with the one of an evaporated metallic top-gate considerably increases the tunability of the p-type structures while the stability of the top-gate can be optimized by the proper choice of a gate insulator oxide. We investigated several quantum point contacts (QPCs)\cite{Wharam1988,Wees1988,Buttiker1990} as test devices patterned by AFM lithography or, alternatively, by electron beam lithography (EBL) and wet chemical etching prior to top-gate deposition. All QPCs were characterized at temperatures between 5 K and 100 mK by four-terminal lock-in measurements of the linear conductance $G$ at a frequency of 31Hz. Here we present representative data obtained on an EBL defined QPC.

Previous studies\cite{Hamilton2008,Rokhinson2006,Danneau2008,Shabani2008,Klochan2006} focused mostly on devices fabricated on (311) oriented deep 2DHGs in Si-doped GaAs, exhibiting an anisotropic in-plane Fermi surface. Here we present data on C-doped (100) samples containing a shallow 2DHG with an in-plane isotropic electronic structure. The QPCs were oriented along the high symmetry
crystallographic directions parallel to the cleaved edges of the wafer. The host heterostructure consists of a 5 nm undoped GaAs cap layer, followed by a 15 nm thick, homogeneously doped layer of AlGaAs separated from the 2DHG by a 25 nm thick, undoped AlGaAs spacer layer.\cite{Reuter1999}

\begin{figure}[t!]
\centering
\includegraphics[width=0.95\columnwidth]{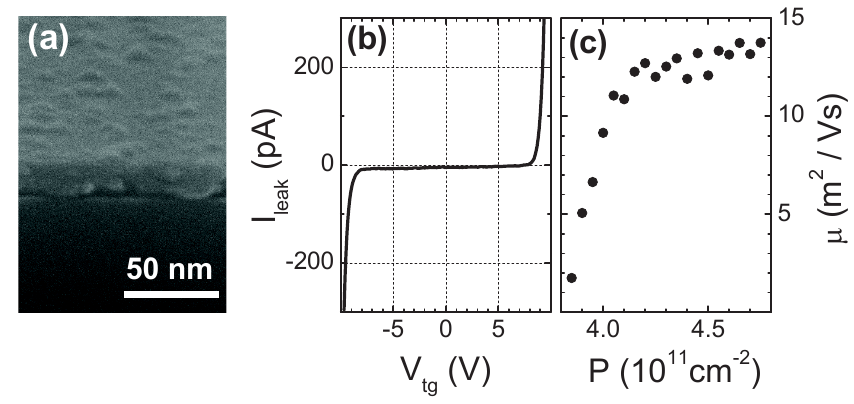}
\caption{(a) SEM image of a freshly cleaved GaAs wafer covered by a 20 nm thick HfO$_{2}$ layer. (b) Breakdown characteristics of the top gate at $T=$ 4.2 K. (c) Hole mobility $\mu$ as a function of the density $P$ in the the 2DHG in the range of -8 V $<V_{\rm tg}<$ 8 V at $T=$ 4.2 K.}
\label{hallbar.fig}
\end{figure}

A 20 nm thick HfO$_{2}$ layer deposited as a gate insulator between the GaAs wafer and the metallic top-gate was grown by atomic layer deposition (ALD) in a Picosun Sunale R-150B system. A linear growth rate of 0.08 nm per pulse cycle has been found at a reaction chamber temperature of 250$^{\rm o}$C and at a tetrakis(ethylmethylamido)hafnium (TEMAH) source temperature of 90$^{\rm o}$C. The structural quality of the resulting polycrystalline HfO$_{2}$ layer have been investigated by scanning electron microscopy (SEM) on freshly cleaved GaAs wafers as illustrated in Fig.~\ref{hallbar.fig}(a). The surface roughness determined by AFM is found to be 4 nm rms, independent of the layer thickness.

\begin{figure}[b!]
\centering
\includegraphics[width=0.9\columnwidth]{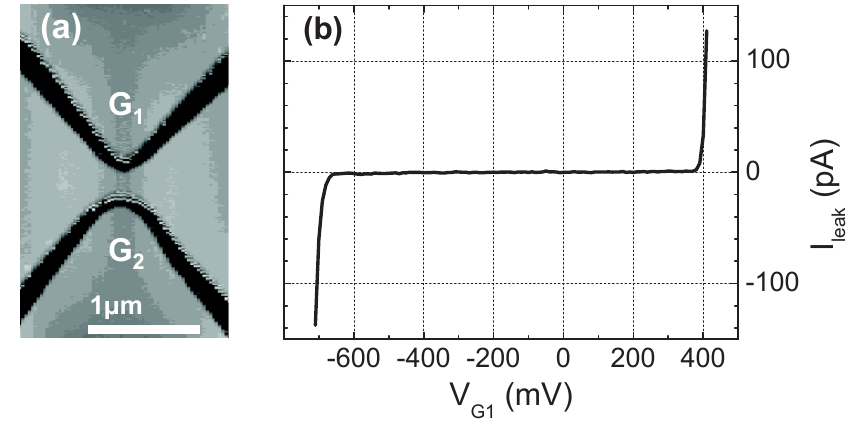}
\caption{(a) AFM micrograph of the 200 nm wide QPC prior to top-gate deposition. The 2DHG is depleted under the etched lines defined by electron-beam lithography. The electronic width of the QPC is controlled by the $G_{1}$ and $G_{2}$ in-plane side gates. (b) Representative breakdown characteristics of the etched lines at $T=$ 4.2 K shown for gate $G_{1}$.}
\label{lines.fig}
\end{figure}

The 10 nm thick Ti sticking layer and the 90 nm thick Au layer of the metallic top-gates were deposited by electron beam evaporation in a Leybold Univex 500 chamber at a base pressure of 10$^{-7}$ mbar. The electronic properties of the oxide layer as well as the tunability of the host 2DHG by the top-gate were tested separately at $T=$ 4.2 K on a 100 $\mu$m wide Hall-bar device defined by conventional optical lithography. The breakdown characteristics of the 20 nm thick oxide layer shows no trace of leakage currents in the -8 V $<V_{\rm tg}<$ 8 V top-gate voltage regime and is reproducible as long as the high voltage leakage current is kept below 0.5 nA as shown in Fig.~\ref{hallbar.fig}(b). The tunability of the hole density $P$ in the 2DHG is, however, limited to about 20-25\% within the leakage free top-gate voltage regime [Fig.~\ref{hallbar.fig}(c)] in contrast to the prediction of a plate capacitor model taking the thicknesses and the dielectric constants of the separating layers into account. This is presumably due to the screening effect of the low mobility charges residing in the doping layer and at the interface between the insulator and the semiconductor. The mobility edge arising from the background potential roughness of the 2DHG is close to the density of the ungated sample $P\approx4\times10^{11}$ cm$^{-2}$.

The structural and electronic properties of the etched lines separating the in-plane gates were individually characterized prior to top-gate deposition. Line widths of 140-170 nm and depths of 15-20 nm provided breakdown voltages on the in-plane gates in the range of a few 100 mV as illustrated in Fig.~\ref{lines.fig}(b) for the EBL defined QPC shown in Fig.~\ref{lines.fig}(a).

\begin{figure}[t!]
\centering
\includegraphics[width=0.95\columnwidth]{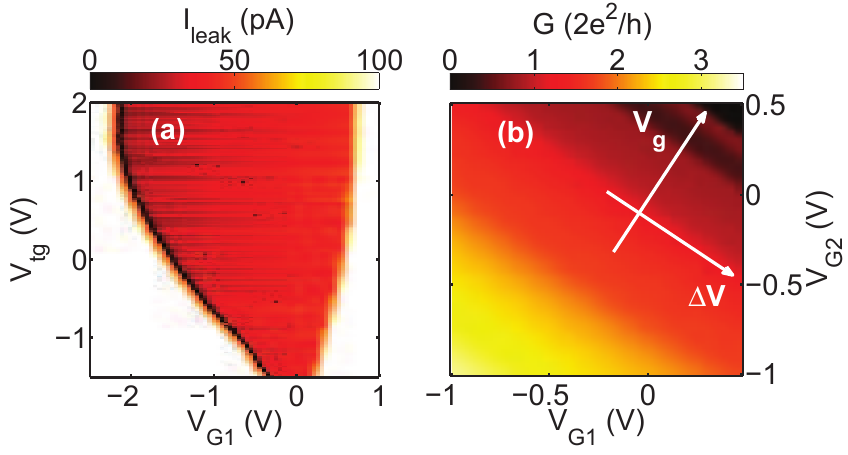}
\caption{(Color online) (a) Leakage current $I_{\rm leak}$ across the in-plane barrier of gate $G_{1}$ as a function of the $V_{\rm tg}$ top-gate voltage and the $V_{\rm G1}$ in-plane gate voltage at $T=$ 100 mK. The colored area shows the magnitude of $I_{\rm leak}$ in the leakage free domain defined as $I_{\rm leak}<$ 0.1 nA. (b) Color map of the QPC conductance $G$ as a function of the $V_{\rm G1}$ and $V_{\rm G2}$ in-plane gate voltages at a representative top-gate voltage value of $V_{\rm tg}=$ 0.5 V. The white arrows indicate the $V_{\rm g}$ and $\Delta V$ symmetric and antisymmetric in-plane gate voltage configurations, respectively (see text for details).}
\label{maps.fig}
\end{figure}

Figure~\ref{maps.fig}(a) shows the absolute value of the leakage current across a separating line in the leakage free gate configuration regime defined as $I_{\rm leak}<$ 0.1 nA when both the $V_{\rm tg}$ top-gate and one of the in-plane gate voltages [shown for $V_{\rm G1}$ in Fig.~\ref{maps.fig}(a)] are varied. The overall broadening of the leakage free domain in $V_{\rm G1}$ with respect to the $I_{\rm leak}(V_{\rm G1})$ trace recorded in the absence of the top-gate [Fig.~\ref{lines.fig}(b)] is explained by the screening effect of the top-gate. The further broadening towards higher $V_{\rm tg}$ reveals the variation of the hole density in the 2DHG, i.e., a decreasing electrochemical potential with respect to the in-plane barriers. The asymmetry in the position of the border lines [also visible in Fig.~\ref{lines.fig}(b)] reflects the asymmetry in the applied in-plane gate biases as one side of the separating lines was always kept at ground. The asymmetry in the curvature of the border lines is attributed to the nonlinearities in the tunneling probability across the barrier under asymmetric in-plane gate biases.

Figure~\ref{maps.fig}(b) displays the conductance of the QPC shown in Fig.~\ref{lines.fig}(a) as its electronic width is tuned by the $G_{1}$ and $G_{2}$ in-plane side gates at a fixed top-gate voltage value of $V_{\rm tg}=$ 0.5 V. The linear combination $V_{\rm g}=\alpha_{1}V_{\rm G1}+\alpha_{2}V_{\rm G2}$ tunes the confining potential in a symmetric fashion, where $-\alpha1/\alpha2$ is the slope of the lines of equal linear conductance on the $V_{\rm G1}$\,--\,$V_{\rm G2}$ plane. Accordingly, the asymmetric gate voltage combination $\Delta V=\alpha_{2}V_{\rm G1}-\alpha_{1}V_{\rm G2}$ leads to a transverse electric field resulting in a lateral shift of the QPC axis while the Fermi energy is kept constant by the leads.

\begin{figure}[t!]
\centering
\includegraphics[width=\columnwidth]{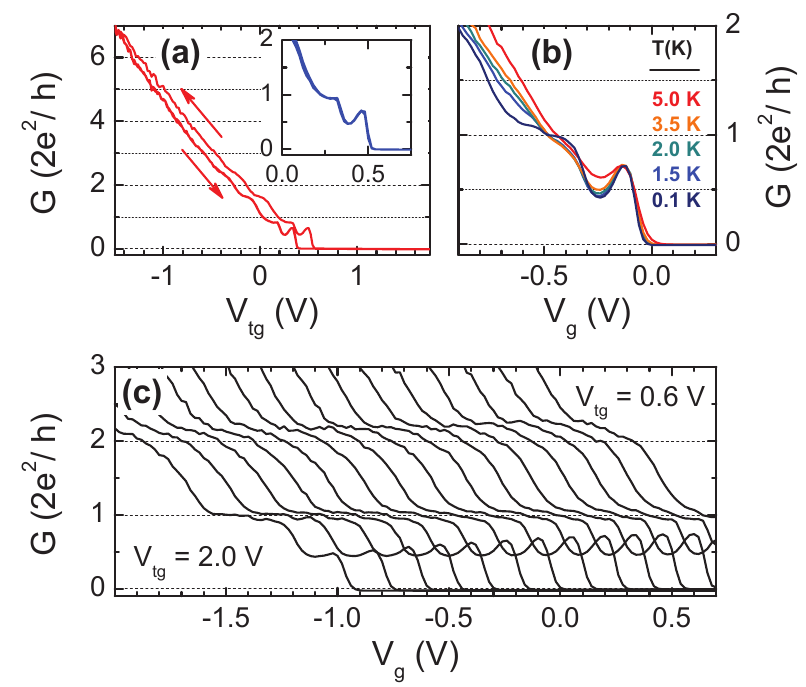}
\caption{(Color online) (a) QPC conductance $G$ as a function of $V_{\rm tg}$ at $T$ = 100 mK while $V_{\rm g}$ is kept at a constant value of 0.5 V. The arrows indicate the direction of the consecutive $V_{\rm tg}$ sweeps. The inset shows a hysteresis-free top-gate sweep in a limited $V_{\rm tg}$ regime. (b) $G$ as a function of $V_{\rm g}$ at various temperatures between $T$ = 5 K and 100 mK at $V_{\rm tg}$ = 1.4 V. (c) $G$ as a function of $V_{\rm g}$ recorded at different top-gate voltages ranging from $V_{\rm tg}$ = 2 V (leftmost trace) to $V_{\rm tg}=$ 0.6 V (rightmost trace).}
\label{qpccond.fig}
\end{figure}

In a simple model of the QPC potential\cite{Buttiker1990} $V_{\rm tg}$ controls the position of the electrochemical potential with respect to the QPC barrier, while $V_{\rm g}$ is expected to set the electronic width of the QPC confinement and thus, also the separation of the one-dimensional subbands. In this picture the onset of the finite conductance occurs as the electrochemical potential approaches the first QPC subband. An increased tunability of the QPCs is expected by having the individual control over both parameters.

The stability of the conductance traces upon different top-gate sweeps is illustrated in Fig.~\ref{qpccond.fig}(a). Sweeping $V_{\rm tg}$ over the extended regime of Fig.~\ref{maps.fig}(a) at a fixed $V_{\rm g}$ value results in a hysteretic behavior due to the charging of the defect sites in the doping as well as in the oxide layer. On the other hand, a top-gate sweep performed in the limited voltage range displayed in the inset of Fig.~\ref{qpccond.fig}(a) yields to hysteresis-free, stable conductance traces. The tunability of the QPC by the in-plane gates at various $V_{\rm tg}$ values is shown in Fig.~\ref{qpccond.fig}(c). It demonstrates that while the overall shape of the individual conductance curves is slightly affected, the range of tunability by means of $V_{\rm g}$ can be optimized by the proper choice of the additional control parameter $V_{\rm tg}$. Moreover, our experience shows that QPCs which were not tunable prior to top-gate deposition due to a large misalignment between the electrochemical potential and the first QPC subband become tunable by changing $V_{\rm tg}$.

At more positive $V_{\rm tg}$ the density in the leads is lower and the opening of the QPC requires more negative voltages on the in-plane gates corresponding to a broader lateral confinement. At the same time, the length of the observed quantized plateaus slightly increase in agreement with the saddle-point potential model.\cite{Buttiker1990} Based on the temperature dependence shown in Fig.~\ref{qpccond.fig}(b) along with further criteria for the associated zero bias anomaly discussed elsewhere\cite{Komijani2010} we attribute the resonant feature on the rise of the $2e^{2}/h$ plateau to a transmission resonance arising from the interaction with an impurity state inside the QPC. The effect of this impurity resonance on the transmission becomes more pronounced towards narrower QPC confinements set by less positive $V_{\rm tg}$ top-gate voltages.

In conclusion, we established a hybrid fabrication method for nanostructures implemented in shallow 2DHGs of GaAs heterostructures utilizing the complementary electrostatic effects of intrinsic in-plane gates and evaporated metallic top-gates. The electric properties of the in-plane barriers as well as of the HfO$_{2}$ top-gate insulator layer were characterized separately. The increased tunability of QPCs fabricated by this method was demonstrated by electrical conductance spectroscopy.

This research was supported by the Swiss National Science Foundation, the German Science Foundation and the German Ministry for Science and Education. A.D.W. and D.R. thank the SFB491, SPP1285 and the BMBF nanoQUIT for financial support. M.C. is grateful to the European Commission for financial support under a Marie Curie Intra European Fellowship.


\end{document}